\documentclass[aps,pre,twocolumn,showpacs,amssymb,floatfix]{revtex4}
\usepackage{amsmath,graphicx,bm}
\newcommand{\tmem}[1]{{\em #1\/}}
\newcommand{\tmop}[1]{\operatorname{#1}}
\newcommand{\bignone}{}
\newcommand{\etal}{{\em et al.~}}
\newcommand{\ie}{{\em i.e.,~}}
\newcommand{\ii}{{\rm i}}
\begin{document}
\title{Critical Frontier of the Triangular Ising Antiferromagnet in a Field}
\author{X. F. Qian~$^{\S}$, M. Wegewijs~$^{\dag}$ and
H. W. J. Bl\"ote~$^{{\ast,\S}}$} 
\affiliation{$^{\S}$Lorentz Institute, Leiden University,
  P.O. Box 9506, 2300 RA Leiden, The Netherlands\\
$^{\dag}$Institut f\"ur Theoretische Physik, RWTH Aachen, 52056 Aachen,
 Germany \\
$^{\ast}$Faculty of Applied Sciences, Delft University of
Technology, P.O. Box 5046, 2600 GA Delft, The Netherlands}

\date{\today}
\begin{abstract}
We study the critical line of the triangular Ising antiferromagnet in
an external magnetic field by means of a finite-size analysis of
results obtained by transfer-matrix  and Monte Carlo techniques.
We compare the shape of the critical line with predictions of two
different theoretical scenarios. Both scenarios, while plausible,
involve assumptions.  The first scenario is  based on the 
generalization of the model to a vertex model, and the assumption that
the exact analytic form of the critical manifold of this vertex model is
determined by the zeroes of an O(2) gauge-invariant polynomial in the
vertex weights. However, it is not possible to fit the coefficients of
such polynomials of orders up to 10, such as to reproduce the numerical
data for the critical points.
The second theoretical prediction is based on the assumption that
a renormalization mapping exists
of the Ising model on the Coulomb gas, and analysis of the resulting
renormalization equations. It leads to
a shape of the critical line that is inconsistent with the first
prediction, but consistent with the numerical data.
\end{abstract}
\maketitle
 
\section { Introduction }\label{int} 
The triangular Ising model with equal nearest-neighbor coupling $K$
in a magnetic field has the reduced Hamiltonian
\begin{equation}
{\mathcal H}/k_{\rm B}T = -  K \sum_{\langle{i,j}\rangle} {s}_{i} {s}_{j}
-H\sum_{{k}} {s}_{k}
\label{Ising}
\end{equation}
where ${s}_{i}=\pm 1$, and $\langle{i,j}\rangle$  indicates summation over
all pairs of nearest-neighbor sites.
According to the exact solution by Houtappel \cite{Hout} of the triangular
Ising model in the absence of a magnetic field, the antiferromagnetic model 
has no phase transition at nonzero temperatures.
The ground state is characterized by the condition that every elementary
triangle contains spins of different signs. This constraint still leaves
a considerable degeneracy, to such an extent that the zero temperature
antiferromagnet has a nonzero entropy.  The ground state appears to
have interesting properties. It is a {\em critical} state as shown by
exact calculations \cite{ST} of the spin-spin correlation function
which appears to decay as a power-law of the distance.
A nonzero temperature $T>0$ destroys the critical state: the correlations
then decay exponentially. However, for sufficiently low $T$, a sufficiently
strong field $H>0$ induces a phase transition to a long-range ordered
state, where the minus spins condense on one of the three sublattices.
As noted by Alexander \cite{SA}, the threefold
symmetry of the ordered phase indicates that the transition belongs to
the three-state Potts universality class. The nature of the transition
was confirmed by Kinzel and Schick \cite{KS}, using phenomenological
scaling \cite{PN} and numerical transfer-matrix calculations; see also
Noh and Kim \cite{NK} and Tamashiro and Salinas \cite{TS}.

\begin{figure}
\includegraphics[angle=0, width=7.5cm]{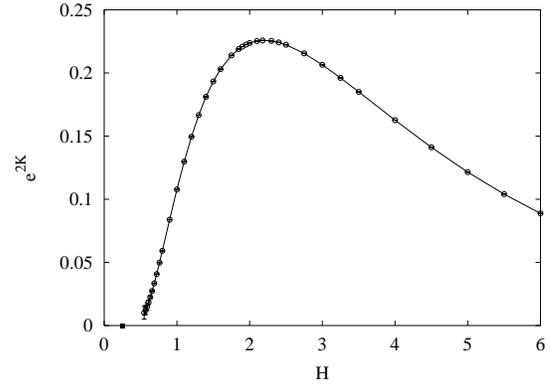}
\caption[xxx]
{Numerical results for the $(H,{\rm e}^{2K})$ phase diagram.
The circles ($\circ$) denote the data points for Ising temperatures $T>0$
and the box ({\tiny $\blacksquare$}) the so-called KT point at $T=0$. }
\label{fig1}
\end{figure}

The critical line covers an infinite range of $K<0$ and $H$. A preview
of our numerical data is given in Fig.~\ref{fig1}.  Since the
phase diagram is symmetric in $H$, we restrict it to $H\geq 0$.
For $K \to -\infty$, $H \to \infty$ while $6K+H$ remains finite, the
model maps \cite{ZR} onto Baxter's hard-hexagon lattice gas of which
the critical exponents are exactly known, and they do indeed fit
the three-state Potts universality class. The asymptotic form 
of the critical line in this lattice-gas (LG) limit is
\begin{equation}
  K_c ( H ) \simeq  - \frac{1}{6} H - \frac{1}{12} \ln \zeta_c \, , 
  \mbox{\hspace{5mm}}
  \zeta_c = \frac{11 + 5 \sqrt{5}}{2} \label{eq:lglimitKH}
\end{equation}
where $\zeta_c$ is the exact critical fugacity calculated by Baxter
\cite{BaxterHH}.

The critical line also extends to $K \to -\infty$ at small fields $|H|$.
The behavior of the critical line in this limit has attracted attention
because of the above peculiar ground-state properties, and the associated
analytical and computational difficulties.
It has been conjectured \cite{KS} that the critical line comes in
vertically in the $\frac{1}{K}$ versus $\frac{H}{K}$ diagram. In other
words, when the Ising temperature goes to zero, also the {\em reduced}
critical field $H_c$ (which includes a factor $1/k_{\rm B}T$) was
supposed to go to zero.
However, Nienhuis \etal \cite{N} provided evidence that $H_c$ instead
approaches a nonzero constant when $K \to -\infty$. This result is 
based on an exact mapping of the zero-temperature Ising model on a
solid-on-solid (SOS) model \cite{BH}. Using renormalization
arguments,  Nienhuis \etal  obtained several critical exponents   
associated with physical fields. It was found that the reduced
magnetic field is {\em irrelevant}: it does not immediately destroy
the critical state at $K = -\infty$.

This renormalization analysis is not rigorous but still convincing.
Several of its predictions agree with exact calculations  \cite{BN}
at $H=0$. The renormalization picture has been extended to include
a nonzero field $H$ and Ising temperature $T$, as well as
next-nearest-neighbor interactions \cite{N}. It predicts that for
$T=0$ the model undergoes an infinite-order transition to a long-range
ordered phase at a finite value of the field $H$. In the SOS language
this is a roughening transition, in the universality class of the
Kosterlitz-Thouless (KT) transition \cite{KT}. The character of this
transition was confirmed \cite{BN,BNH} and located at $H_{\rm KT}=0.266(10)$
by means of transfer-matrix calculations and phenomenological
renormalization \cite{BN}. The associated finite-size-scaling
analysis is problematic because of slow convergence due to logarithmic
corrections at the KT transition point. Such corrections are possibly
a reason why an analysis by de Queiroz \etal \cite{QP}, without such
corrections, yielded a result that is not fully consistent, namely
$H_{\rm KT}=0.211(7)$ (for the correct interpretation of this result
it is essential that the field $H$ used in Fig.~1 and Table~I of
Ref.~\cite{QP} does {\em not} contain a factor $1/T$ \cite{Qp}).

The estimated critical field $H_{\rm KT}$ at $T=0$ appears to be much 
smaller than estimates obtained at $T>0$. The question thus arises whether
the Potts critical line for $T>0$ connects to the KT point at $T=0$.
It is noteworthy that the renormalization scenario given in Ref.~\cite{N},
which includes next-nearest-neighbor interactions $K_{\rm nnn}$, implies
that the line of phase transitions limiting the ordered phase in the
$(K_{\rm nnn},T)$ plane does \emph{not} connect to the transition line
in the $(K_{\rm nnn},H)$ plane. Thus one may ask the same question for
the  $(K_{\rm nnn},H)$ and the $(H,T)$ plane. An answer to this question
is provided by renormalization arguments presented in Section~\ref{sec3}.
This approach also predicts the analytical form of the Potts critical line
for $T \to 0$ while $H$ remains finite.\\

A different approach to find the shape of the critical line of an
antiferromagnetic
Ising model in a field was formulated by Wu \cite{FYWu} who noted that
these models can be mapped on vertex models, and that these vertex models
have symmetry properties that impose restrictions on the analytic form
of their critical manifolds. He also noted that the critical manifolds
of the exactly solved vertex models are determined by the zeroes of 
homogeneous polynomials in the vertex weights that are invariant under
the symmetry group of the model.  On the basis of the
assumption that the latter form of the critical subspace also applies
to vertex models that are equivalent with antiferromagnetic Ising models
in a field \cite{WuWu}, one may thus attempt to solve for the unknown
independent coefficients of the homogeneous polynomial, the number of 
which is dramatically reduced by symmetry restrictions.
In actual applications, the number of equations is still not enough
to solve all unknown coefficients, and additional numerical input is
required, for instance from a numerical transfer-matrix analysis.

This approach is more ambitious than the renormalization analysis in
the sense that its aim is to describe the {\em whole} critical manifold.
It has been applied to the Ising antiferromagnets on the honeycomb
lattice \cite{WWB} and on the square lattice \cite{BW}.
The transfer-matrix data, with accuracies in the order of $10^{-10}$,
could be successfully described by such invariant polynomials of
relatively low order.  Nevertheless one may remark
that these analyses did not provide solid evidence for the {\em exact}
form of the critical line of the Ising antiferromagnet.

Application of this approach to the triangular Ising model leads to some
additional complications. First, the topology of the phase diagram is less
simple,  which relates to the fact that the lattice is not bipartite.
Second, the 3-state Potts character of the critical line implies
that corrections-to-scaling  converge less well in comparison with the
Ising case, so that it is not feasible to reach the same 
degree of numerical accuracy.

In this paper we compare the results of both theoretical approaches to
our numerical data for the triangular Ising antiferromagnet.
 In Sec.~\ref{sec2} we formulate the invariant-polynomial scenario and
 derive an exact restriction on the critical line
 which must hold if this line is analytic in the KT point. 
 We explicitly construct invariant polynomials of arbitrary order in the
 Ising vertex weights whose roots exhibit this behavior.
 A summary of the Coulomb gas scenario and an analysis of the 
renormalization-flow equations follows in Sec.~\ref{sec3}. The analytic
forms of the critical lines predicted by these two scenarios appear 
to be mutually {\em inconsistent} for $T \to 0$ at finite $H$.
In Sec.~\ref{sec4} we outline our transfer-matrix construction and
present accurate results for the critical points. This section also
includes a Monte Carlo analysis of the critical amplitudes. An analysis
and a discussion of these results is given in Sec.~\ref{sec5}.
Finally, we draw our conclusions in Sec.~\ref{con}.

\section{O(2) Invariant polynomials in the vertex weights}
\label{sec2}
The mapping of the triangular Ising model on the 64-vertex model
involves the introduction of bond variables $b_{ij}=0$ or 1 between
nearest-neighbor spins and summation over the Ising variables \cite{FYWu}.
Since the bond variables are independent, there are $2^6=64$ distinct
vertices. But these turn out to have only 7 distinct weights
\begin{equation}
  W_{s_1 \ldots s_6} = W_s = h^{s \tmop{mod} 2} z^{s / 2} \label{eq:wising}
\end{equation}
where $h = \tanh ( H )$ and $ z = \tanh ( K )$. These weights
are 'symmetric' \ie depend only on the number of covered
bonds $\sum_{j}^{\rm nn} b_{ij} $ connecting a vertex $i$ to its 6
nearest neighbors $j$. The weights can take imaginary
($s =$odd, $K < 0$) or real (otherwise) values.
The partition function of an $N$-site system is a
homogeneous polynomial of order $N$ in the weights:
\begin{equation}
   Z_{\text{vertex}} \equiv \sum_{\{ b_{i j} \}} \prod_k W_{s_1 ( k )
  \ldots s_6 ( k )} \label{eq:zvertex}
\end{equation}
Both pertinent indices of neighboring vertices $i$ and $j$ in this product
are set equal to the bond variable $b_{i j} = 0, 1$. The summation
runs over all possible configurations of the bond variables. The 
partition function of the Ising model differs from Eq.~(\ref{eq:zvertex})
only by a multiplicative factor which is non-singular for finite $H$
and $ K$:
\begin{displaymath}
  Z_{\text{Ising}} ( H, K ) =\mbox{\hspace{50mm}}
\end{displaymath}
\vspace{-6mm}
\begin{equation}
2 \cosh (H)^N \cosh ( K )^{\frac{3}{2}N} \times
  Z_{\text{vertex}} ( \{ W_s ( h, z ) \} )
\end{equation}
A crucial property of Eq.~(\ref{eq:zvertex}) is that the summation over one of
the bond variables $b_{i j}$ is invariant with respect to any O(2)
transformation
\begin{equation}
  R ( \theta, \varepsilon ) = \left(\begin{array}{cc}
    \cos \theta & - \varepsilon \sin \theta\\
    \sin \theta & \varepsilon \cos \theta
  \end{array}\right)
\end{equation}
with respect to the indices of the connected vertices. Here,
$\det ( R ) = \varepsilon=\pm 1$ distinguishes the SO(2) subgroup of
proper rotations ($\varepsilon = + 1$) from the improper transformations
($\varepsilon = - 1$) which also include a reflection.
Application of this transformation to all bonds connecting neighboring
vertices (assuming periodic boundary conditions) leads to a partition
sum of the same form but with new weights
\begin{equation}
  W_{s_1' \ldots s_6'}' = R_{s_1' s_1} ( \theta, \varepsilon ) \ldots R_{s_6'
  s_6} ( \theta, \varepsilon ) W_{s_1 \ldots s_6} \label{eq:wtrans}
\end{equation}
where we use the dummy summation convention $s_i = 0, 1$. This gauge
transformation preserves the symmetry mentioned under Eq.~(\ref{eq:wising}).
However, only special O(2) transformations preserve the Ising weight
parametrization  expressed by the right-hand side of Eq.~(\ref{eq:wising}).
A trivial example is the reflection $\theta = 0,
\varepsilon = - 1$ effecting $W_s ( h, z ) \rightarrow W_s ( - h, z )$. This
corresponds to an external field inversion $H \rightarrow - H$. Below we will
first discuss another less trivial transformation which also leads to weights
of the Ising form (\ref{eq:wising}) up to a common factor.
This will have consequences for the
asymptotic behavior of the critical curve $h_c ( z )$. In order to
investigate the assumption that the critical curve is a root of an O(2)
invariant homogeneous polynomial in the vertex weights, we explicitly
construct these polynomials of arbitrary order. Conclusions about their
compatibility with our numerical data will be drawn in Section~\ref{sec5}.

\subsection{Dual transformation}

The O(2) transformation $\theta = \frac{\pi}{2}, \varepsilon = \mp 1$ changes
the weights as
$W_s (h,z) \rightarrow z^3 W_s \left( \pm \tmop{sign} (z) h, 1/z \right)$.
This connects (up to a non-singular factor) the Ising
weights (\ref{eq:wising}) for physical values of $|h|, |z| \leqslant 1$ to
weights of the same form but with unphysical values $|h|, |z| \geqslant 1$
corresponding to complex fields $H \pm \ii \pi / 2, K \pm \ii \pi / 2$.
Extending the vertex model with weights (\ref{eq:wising}) to all real
values of $h, z$ the antiferromagnetic ($z < 0$) critical curve of this model
$h_c ( z )$ has a physical and non-physical branch which are connected by the
dual transformation
\begin{equation}
\text{$( h, z ) \rightarrow ( h, 1 / z )$} \label{eq:dual} 
\end{equation}
and which is separated by the self-dual LG and KT points at $z = - 1$.
This is a rigorous result of seemingly little use. However, we will now
demonstrate that, combined with the {\tmem{assumption}} that the critical
curve is analytical for $z = - 1$, this severely constrains the shape of the
antiferromagnetic branches of the critical curve for {\tmem{physical}} values
of $h, z$. Consider the derivatives of a branch of the critical line
$h_c(z)$ at $z = - 1$, assuming it is analytical there. Differentiating the
constraint $h_c ( z ) = h_c ( 1 / z )$ $n$ times and recursively solving for
the derivatives $h_c^{( k )} = \partial^k h_c / \partial z^k |_{z = - 1}$ up
to order $k = 1, \ldots, n$ one finds that each {\tmem{odd}} order derivative
can be expressed as a linear combination of lower {\tmem{even}} order
derivatives with integral coefficients. The expansion of $h_c ( z )$ must take
the form
\begin{equation}
  h_c = h_c^{( 0 )} + \frac{1}{2 m!} h_c^{( 2 m )} \left[ ( 1 + z )^{2 m} + m
  ( 1 + z )^{2 m + 1} \right] + \ldots \label{eq:hcexp}
\end{equation}
Here $m \geqslant 1$ is some integral number \ie  the first non-vanishing term
is of even order. A notable feature of Eq.~(\ref{eq:hcexp}) is that, if 
the critical line is analytical at $z = - 1$, it {\tmem{must}} satisfy
${\rm d}h_c /{\rm d}z=0$. We now consider the implications for the KT
and LG asymptotic lines. First, the LG asymptotic relation
Eq.~(\ref{eq:lglimitKH}) expressed in the variables $h, z$ reads:
\begin{equation}
  \frac{1 + h}{1 - h} = \frac{1}{\zeta_c} \left( \frac{1 - z}{1 + z} \right)^6
  \label{eq:lgasymp}
\end{equation}
which indicates analyticity of the exact form of the LG branch of the
critical line at $z = - 1$.  The first {\tmem{two}} non-vanishing terms
 in the resulting expansion
\begin{equation}
  h_c ( z ) = 1 - \frac{\zeta_c}{2^5} ( ( 1 + z )^6 + 3 ( 1 + z )^7 + \ldots )
\end{equation}
 are the same as in the expansion of Eq.~(\ref{eq:lgasymp}) with $m=6$.
We note that the LG asymptotic curve Eq.~(\ref{eq:lgasymp}) is invariant
under $( h, z ) \rightarrow ( h, 1 / z )$ 
up to all orders. This is 
due to the dual symmetry of the corresponding vertex model.
Secondly, the approach
of the critical curve to the KT asymptotic line must also be of the form 
(\ref{eq:hcexp}). However, the integer value $\text{$m \geqslant 1$}$ in this
case is unknown. Expressed in the physical variables $H, K$, we find a
{\tmem{logarithmic}} divergence:
\begin{equation}
  K ( H ) = \frac{1}{4 m} \ln ( H_c - H_{\tmop{KT}} ) + \tmop{const} .
  \label{eq:ktasympto2}
\end{equation}
This is the central result of this section.  We emphasize that it is
based on the assumption that the KT branch of critical line is analytic
at $z = - 1$.  Roots of  O(2) invariant polynomial equations, which we
construct explicitly below, all have this property and in the general
case $m = 1$.

\subsection{O(2) invariant polynomials in the Ising weights}

Now we explicitly construct homogeneous polynomial equations in the Ising
weights which incorporate all the constraints imposed by the O(2) gauge
symmetry (including those discussed above). Following Perk \etal
{\cite{bib:perk}} we change to the eigenbasis of the SO(2)
subgroup of rotations ($\varepsilon = + 1$) via the 2$\times$2 matrix
$\alpha_{s_i \sigma_i} =  \frac{1}{2} ( - \ii \sigma_i )^{s_i}$:
\begin{equation}
  W_{s_1 \ldots s_6} = \alpha_{s_1 \sigma_1} \ldots \alpha_{s_6 \sigma_6}
  A_{\sigma_1 \ldots \sigma_6}
\end{equation}
Note that $s_i = 0, 1$ but $\sigma_i = \pm 1$. In this basis the
transformation Eq.~(\ref{eq:wtrans}) takes the simple explicit form
($\varepsilon = \pm 1$):
\begin{equation}
  A_{\sigma_1 \ldots \sigma_6}' = {\rm e}^{\ii \theta ( \sigma_1 + \ldots + \sigma_6
  )} A_{\varepsilon \sigma_1 \ldots \varepsilon \sigma_6}
\end{equation}
For the symmetric vertex model, the 7 components transform as
$A_{\sigma}' = {\rm e}^{i \theta \sigma} A_{\varepsilon \sigma}$, where
$\sigma = \sigma_1 + \ldots + \sigma_6 = 0, \pm 2, \pm 4, \pm 6$. In this 
basis the Ising weights are complex:
\begin{displaymath}
  A_{2 k} = ( z + 1 )^{3 - k}  \hspace{50mm}
\end{displaymath}
\vspace{-6mm}
\begin{equation}
\left[ \sum_{m = 0}^k \binom{2 k}{2 m} ( - z )^m
  + \ii h \sqrt{z} \sum_{m = 0}^{k - 1} \binom{2 k}{2 m + 1} ( - z )^m \bignone
  \right] \label{eq:Aising}
\end{equation}
where $k = 0, 1, 2, 3$ and $A_{- 2 k} = A_{2 k}^{\ast}$.  Invariant
polynomials which transform with parity $\pm 1$ (\ie  $I_{\pm} \rightarrow \pm
I_{\pm}$) have the simple form $I_{\pm} = \frac{1}{2} ( \prod_{\sigma}
A_{\sigma}^{n_{\sigma}} \pm \prod_{\sigma} A_{\sigma}^{n_{- \sigma}} )$ (or a
linear combination of these) with exponents which satisfy $\sum_{\sigma}
\sigma n_{\sigma} = 0$. A minimal finite set of such exponents can be found
which generate all other solutions by linear combination with {\tmem{integer}}
coefficients. This implies that any O(2) invariant $I_{\pm}$ can be generated
as a {\tmem{polynomial}} function of a minimal set of so-called
 fundamental invariants. These 14 polynomials have been constructed in
{\cite{bib:perk}}.  The crucial point is to eliminate all  dependencies due
to polynomial relations between the fundamental invariants (called syzygies),
and further dependencies introduced by the parametrization (\ref{eq:Aising}).
To generate invariants of parity $+ 1$ we need to retain only 4 fundamental
O(2) invariants of parity $+ 1$ which can be compactly be written in variables
$s = 1 - h^2 = 1 / \cosh^2 H$
and $ u = 1 + z = {\rm e}^K / \cosh K$
\begin{equation}
  \begin{array}{lllll}
    I_0 & = & A_0 & = & u^3\\
    I_k & = & A_{2 k} A_{- 2 k} & = & u^6 + u^{6 - 2 k} \Omega_k ( u ) s
  \end{array}
\end{equation}
where
\begin{equation}
  \begin{array}{lll}
    \Omega_1 ( u ) & = & 4 ( 1 - u )\\
    \Omega_2 ( u ) & = & 4 ( 1 - u ) ( 2 u - 4 )^2\\
    \Omega_3 ( u ) & = & 4 ( 1 - u ) ( 3 u - 4 )^2 ( u - 4 )^2
  \end{array} \label{eq:omega}
\end{equation}
The most general invariant polynomial of even order $2 e \geqslant 2$ and
parity $+ 1$ in the Ising weights is
\begin{equation}
  f_{2 e} = \sum_{j = 0 \bignone}^e I_0^{2 ( e - j )} \left[ \sum_{l = 0}^j
  c_l^j I_2^l I_3^{j - l} + \sum_{l = 1}^j c_{j + l}^j I_1^l I_2^{j - l}
  \right] \label{eq:f2ec}
\end{equation}
Cross terms of $I_1$ and $I_3$ have been eliminated using a polynomial
relation. The $( e + 1 )^2$ coefficients in Eq.~(\ref{eq:f2ec}) correspond
1-to-1 to an invariant expression in the Ising weights of order $2 e$. To
exclude a trivial factorization to an even order polynomial, \ie  $f_{2 e} =
I_0^2 f_{2 ( e - 1 )}$, we require $c_0^0 \neq 0$. Polynomial invariants of
{\tmem{odd}} order and parity $+ 1$ can be shown to factorize trivially,
$f_{2 e + 1} = I_0 f_{2 e}$, and thus need not be considered further. Finally,
polynomial invariants of any order and parity
$- 1$ can be discarded
also. We find that for $z = - 1$ such polynomials have no other root
than $h = \pm 1$ \ie  the KT point cannot be described. The expansion of $f_{2
e}$ in Eq.~(\ref{eq:f2ec}) is not well suited to impose restrictions on the
coefficients from known properties of the critical curve of the triangular
lattice Ising model. A more convenient but equivalent expansion is obtained by
replacing $I_k \rightarrow I_k - I_0^2, k = 1, 2, 3$ and $c_l^j \rightarrow
\kappa_l^j$  in Eq.~(\ref{eq:f2ec}). This gives the final explicit form
\begin{displaymath}
  f_{2 e} = \sum_{j = 0}^e 
 \left[ \sum_{l = 0}^j \kappa_l^j \Omega_2^l
  \Omega_3^{j - l} u^{2 l} +   \right. \hspace{25mm}
\end{displaymath}
\vspace{-6mm}
\begin{equation}
 \hspace{25mm}  \left.
\sum_{l = 1}^j \kappa_{j + l}^j \Omega_1^l
  \Omega_2^{j - l} u^{2 l + 2 j} \right] \bignone u^{6 ( e - j )} s^j
  \label{eq:f2ek}
\end{equation}
Since $\Omega_k ( u ) \rightarrow 4^{2 k - 1}$ in the KT 
 ($u \rightarrow 0$, $s = 1 - h_{\tmop{KT}}^2 \neq 0$)
and LG ($s, u \rightarrow 0$) limit we can
easily solve $f_{2 e} ( s, u ) = 0$ for the asymptotic relations between $s$
and $u$ in each limit. We find that consistent polynomials must satisfy
$\kappa_0^0 = - 4^3 \zeta_c \kappa_0^1$  where $\zeta_c$ is given by
Eq.~(\ref{eq:lglimitKH}) (LG) and
$\kappa^e_0 = \kappa^e_1 = \kappa^e_2 = 0$ (KT).
This already excludes $f_2 = 0$ as a candidate. We thus have $( e + 1 )^2 - 5
= 4, 11, 20, 31$ independent coefficients in a polynomial of order $2 e = 4,
6, 8, 10$, respectively. The asymptotic value of $s$ in the KT limit is
determined by $s_{\tmop{KT}} = 4 \kappa_0^{e - 1} / \kappa^e_3$. This may be
used to either extract this value after a fitting procedure or as an extra
constraint. One can verify from the general form Eq.~(\ref{eq:f2ek}) that
the approach to the KT value indeed takes the form Eq.~(\ref{eq:ktasympto2})
with $m=1$.  This is dictated by the dual transformation property
Eq.~(\ref{eq:dual}) combined with the analyticity of the branches determined
by $f_{2 e} = 0$. Higher integer values for $m$ are also possible but require
certain coefficients in Eq.~(\ref{eq:f2ek}) to be strictly zero, which is not
supported by the numerical data.

\section{Renormalization analysis}
\label{sec3}
\subsection{Mapping on the Gaussian model }
At zero temperature, the three spins of
each elementary triangle cannot have the same sign. Thus each triangle
has two bonds between antiparallel spins and one bond between parallel spins.
When all bonds between parallel spins are erased, one obtains a lattice
tiling with rhombi. This tiling can also be interpreted as a stack of
cubes viewed from the (1,1,1) direction. Thus the zero-temperature
antiferromagnetic triangular Ising model is equivalent with a
solid-on-solid (SOS) model \cite{BH}. The SOS model consists
of height variables $h_i$ where $i$ denotes the lattice site. The height
variables assume integer values satisfying $h_i \; {\rm mod}\; 3=c_i$,
where $c_i=0$, 1 or 2 denotes the sublattice of site $i$.
Apart from an infinite constant, the Hamiltonian  becomes
\begin{displaymath}
{\mathcal H}/k_{\rm B}T =  K_{\infty} \sum_{\langle{i,j}\rangle}
(1-\delta_{|h_i-h_j|,1})(1-\delta_{|h_i-h_j|,2})
\end{displaymath}
\begin{equation}
 -H \sum_{k}(2 \delta_{(h_k \; {\rm mod} \; 2,0)}-1)
\label{SOS}
\end{equation}
where we set $K_{\infty} \to \infty$ so that the product 
$(1-\delta_{|h_i-h_j|,1})(1-\delta_{|h_i-h_j|,2})$ restricts the height
differences between nearest-neighbor sites to $1$ or $2$. The Kronecker
delta in the second term counts the numbers of $+$ spins.

The equivalence with the Ising model (\ref{Ising}) makes it possible
to express the height-height correlation function \cite{N}
\begin{equation}
g(r) = \langle [(h_{r}-h_{0})-\langle h_{0}-h_{r}
\rangle ]^2 \rangle
\label{HH}
\end{equation}
in terms of Ising correlations. {}From the exact results
for $K=-\infty$, $H=0$, it follows \cite{N} that
\begin{equation}
g(r) \simeq \frac{9}{\pi^2} \ln(r)+{\rm const}
\label{HHC}
\end{equation}
where $r$ is the distance between the correlated sites. This result is
very useful in the context of a renormalization mapping 
by Nienhuis \etal  \cite{N} on the Gaussian model
with Hamiltonian
\begin{equation}
-{\mathcal H}/k_{\rm B}T=
\frac{2\pi}{T_{\rm R}} \sum_{\langle{i,j}\rangle}(h_i-h_j)^2
+\sum_{p} S_{p} \sum_{i}\cos \frac{2\pi h_i}{p}
\label{GS}
\end{equation}
where the second summation contains so called spin-wave perturbations
of the Gaussian model, \ie a periodic potential acting on the Gaussian
height variables. A term with $p=1$ originates from the discreteness
of the height variables in Eq.~(\ref{SOS}). A nonzero magnetic field $H$
favors triangles with only one minus spin; in the SOS model this leads
to an energy alternation between even and odd heights. This maps on
a spin-wave perturbation with $p=2$ in the Gaussian model. 

The mapping of Eq.~(\ref{SOS}) to  Eq.~(\ref{GS}) is not exact, 
so that the renormalized temperature $T_{\rm R}$ is
in principle unknown. However, the height-height correlation function
of the Gaussian model is known to depend on $T_{\rm R}$ as
\begin{equation}
g(r) \simeq \frac{T_{\rm R}}{2\pi^2} \ln(r)
\label{Gaussian}
\end{equation}
Comparison with Eq.~(\ref{HHC}) shows that  
\begin{equation}
T_{\rm R}=18
\label{Gaustmp}
\end{equation}
 for $H=0$.
Once  $T_{\rm R}$ is known, several quantities of interest can be
calculated for the Gaussian model. These quantities include the scaling
dimensions associated with the so-called spin-wave and vortex
perturbations in the Gaussian model.
Since an elementary excitation of the Ising model, \ie a triangle with
three equal Ising spins, leads to an SOS height mismatch of 6 units, the
Ising temperature field $t_{{\rm I}}$ is represented by the fugacity 
$V_q$ of the $q=\pm 6$ vortices. 
On the basis of the known results for the scaling dimensions of $S_{p}$
and $V_q$ in the Gaussian model \cite{JK}, a number of properties of
the triangular Ising model, including a part of the phase diagram
extended in the direction of $T_{\rm R}$, have been derived \cite{N}.
In this work we make use of the language of the Coulomb gas formulation
\cite{BNDL} to express the relevant scaling dimensions. The appropriate
parameters are the renormalized coupling constant  $g_{\rm R}$, and
electric charges $e$ and magnetic charges $m$. Their relation with
the parameters of the Gaussian model can be expressed as
\begin{eqnarray}
g_{\rm R} &=&   36/T_{\rm R}       \nonumber    \\
e         &=&   6/p                \label{defcg}\\
m         &=&   q/6                \nonumber
\end{eqnarray}
In this language, the scaling dimensions $X_{e,m}$ associated with 
the activity of charges $(e,m)$  are
\begin{equation}
X_{e,m}       =\frac{e^2}{2 g_{\rm R}} + \frac{g_{\rm R}m^2}{2}
\end{equation}
{}From Eq.~(\ref{Gaustmp}) we see that $g_{\rm R}=2$ for $H=0$.
Since the Ising temperature field is associated with magnetic
charges $\pm 1$, the Ising temperature renormalization exponent
$y_{t_{\rm I}}=2-X_{0,1} = 1$ is relevant.
The system is thus disordered for all $T>0$. 
However, the exponent  $y_h=2-X_{3,0}=-1/4$  associated with the
uniform magnetic field $H$ is irrelevant.  Thus, at $T=0$ the system
remains critical for a certain range of the Ising field $H$. In the
renormalization scenario outlined in Ref.~\cite{N} a phase transition
to the long-range ordered  phase occurs when $H$ grows large enough.
In the SOS language this ordered phase is flat, and transition is of
the `roughening' type, by duality related to the Kosterlitz-Thouless
phase transition \cite{KT}. 
  
\subsection{Renormalization flow }
\label{rgflow}
For an analysis of the renormalization flow at nonzero magnetic field
$H$ it is necessary to include the Coulomb gas coupling $g_{\rm R}$
because the field, although irrelevant at  $g_{\rm R}=2$, tends to
suppress height differences, \ie  to increase $g_{\rm R}$. As deduced in
Ref.~\cite{N} such an effect can also be realized without
breaking the Ising symmetry, by the introduction of ferromagnetic
next-nearest-neighbor interactions into the model (\ref{Ising}). Although
we restrict our numerical investigation to the 2-parameter model
(\ref{Ising}), the renormalization analysis still requires a set of 3
nonlinear scaling fields, which are chosen as
\begin{eqnarray}
t         &=&   \frac{9}{2 g_{\rm R}} -2                   \nonumber    \\
h         &=&   \alpha_1 H +\alpha_3 H^3 + \cdots  \label{defp} \\
t_{\rm I} &=&   {\rm e}^{2K} + \cdots              \nonumber
\end{eqnarray}
The constant  $\alpha_1$ determines the scale of $t$ which remains to
be determined. Apart from that, the expansion coefficients $\alpha_j$
are in principle unknown. The Ising temperature field $t_{\rm I}$ is
in lowest order chosen as the Boltzmann factor of an elementary Ising
excitation, \ie  a triangle with three equal spins. It thus describes
the activity of the magnetic charges $m=\pm 1$ in the Coulomb gas.
In order to describe the renormalization of these parameters in the
immediate vicinity of the fixed line $h=t_{\rm I}=0$, the following
renormalization exponents apply 
\begin{eqnarray}
y_t          &=&   0                         \nonumber    \\
y_h          &=&   2 - \frac{9}{2 g_{\rm R}} \label{ydef} \\
y_{t_{\rm I}}&=&   2-\frac{g_{\rm R}}{2}     \nonumber
\end{eqnarray}
First we address the special case $T=t_{\rm I}=0$. Because of the
marginality of
$t$ we add a nonlinear term in the flow for $t$. In differential form,
the equations become
\begin{equation}
\frac{{\rm d}h(l)}{{\rm d}l} =-ht\, , \mbox{\hspace{10mm}}
\frac{{\rm d}t(l)}{{\rm d}l} =-h^2
\label{dif2p}
\end{equation}
where $l$ parametrizes the rescaling factor $b$ as $b=\exp(l)$. In 
principle one has an unknown amplitude in the second equation, but we
have disposed of it by a proper choice of $\alpha_1$. Thus, the scale
of $t$ is set such as to simplify the renormalization equations to
(\ref{dif2p}).  The equation for $h$ follows from
the usual form $h'=b^{y_h}h$ after substitution of $b$ in terms of $l$,
of $y_h$ using Eq.~(\ref{ydef}), and $g_{\rm R}$ in terms of $t$. The
sign in the equation for $t$ follows because $h$ suppresses the height
differences in the SOS language. The flow equations (\ref{dif2p})are
equivalent to those describing the Kosterlitz-Thouless \cite {KT} and
roughening transitions.  Elimination of $l$ from Eq.~(\ref{dif2p}) and
integration yields the trajectory in the $h,t$ plane as
\begin{equation}
h^2=t^2+c^2
\label{tra2p}
\end{equation}
where the constant $c$ follows from the initial conditions which are
chosen as $h(0)=1/4+\delta h$, $t(0)=1/4$. The physical motivation
of this choice is that the KT transition line obeys $h=t$, so that
we select a point at a distance $\delta h$ to the KT point of the
nearest-neighbor model Eq.~(\ref{Ising}).  For small $\delta h$ one
finds $c^2 \approx \delta h/2$. Elimination of $h$ in Eq.~(\ref{dif2p})
leads to
\begin{equation}
\frac{{\rm d}t(l)}{{\rm d}l} = -t^2 - c^2
\label{dift2}
\end{equation}
Integration, substitution of the initial conditions and some 
rearrangement lead to the renormalization flow for $t_{\rm I}=0$ as
\begin{equation}
t= \frac{\sqrt{\delta h/2}}{\tan(\sqrt{8 \delta h}+l \sqrt{\delta h/2})}
\label{tfunl}
\end{equation}

Next we introduce a nonzero Ising temperature $T$ so that also $t_{\rm I}>0$.
It seems reasonable to assume that the renormalization flow of $h$ and $t$
is not seriously affected for small $t_{\rm I}$.
For simplicity we make a stronger assumption, namely that the flow of
$h$ and $t$ is independent of that of $t_{\rm I}$.
We first focus on the question whether the 3-state Potts critical line
in the $(H,T)$ plane extends to the KT transition at $T=0$, or, in other
words, whether there are points in the immediate vicinity of
$(H=H_{\rm KT},T=0)$ that flow towards a region where we can be confident
that a Potts-like transition occurs. For small $\delta h/2$ and $t_{\rm I}$
the first part of the path will be governed by the KT fixed point. We
assume that the flow will bring the system to a boundary which separates
the regions governed by the KT and Potts fixed points. This boundary
is obviously not determined in a quantitative sense, but this does not 
matter for the present scaling argument. Let it be sufficient to define
this region by requiring that $h$ and $t_{\rm I}$ reach values of order 1.

We search for this region by choosing the (somewhat arbitrary) renormalized
temperature $g_{\rm R}=3$ where we have evidence \cite{XH1} that the Potts
transition connects to the neighborhood of $T=H=0$.
The shape of the critical line at $g_{\rm R}=3$ is determined by the
flow equations for $t_{\rm I}$ and $h$, namely
\begin{eqnarray}
t_{\rm I}(b) &=&   b^{y_{t_{\rm I}}}t_{\rm I} \nonumber    \\
h(b)         &=&   b^{y_h} h                  \label{thr }
\end{eqnarray}
where one may take $y_{t_{\rm I}}=y_h=1/2$ as long as $t_{\rm I}$ and
$h$ are small so that the change of $g_{\rm R}$ can be neglected. Then,
$t_{\rm I}(b)/h(b)$ is constant along a flow line and can be chosen such
that the model is critical, say for 
\begin{equation}
t_{\rm I}(b)/h(b)=\beta\, .
\label{tovb}
\end{equation}
where $\beta$ is a constant of order 1.  For larger values
of $t_{\rm I}$ and $h$ the relation will no longer be linear but
it is reasonable to expect, and in agreement with numerical
results \cite{Landau,XH1}, that there is a fair range where $\beta$ is
still of order 1.

\begin{figure}
\includegraphics[angle=0, width=7.5cm]{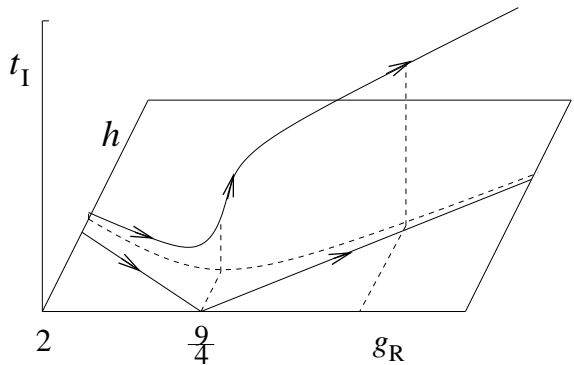}
\caption[xxx]
{Sketch of the renormalization flow in the
parameter space of the renormalized coupling $g_{\rm R}$, the scaling
field $h$ and the Ising temperature field $t_{\rm I}$. The flow of
$g_{\rm R}$ and $h$ is anomalously slow near the point $g_{\rm R}=9/4$,
$h=0$: most of the growth of  $t_{\rm I}$ occurs here.
}
\label{fig2}
\end{figure}

Thus we consider a point  $h(0)=1/4+\delta h$, $t(0)=1/4$,
$t_{\rm I}(0) = \delta t_{\rm I}$ in the vicinity of the KT point and
apply a transformation such  that the system flows to $g_{\rm R}=3$ or
$t=-1/2$. According to Eq.~(\ref{tfunl}) the scale factor
of this transformation is $b=\exp(l) = \exp(\pi \sqrt{2 /\delta h})$.
It follows from Eq.~(\ref{tfunl}) that (for small $\delta h$ and 
$\delta t_{\rm I}$) the system is located near the KT fixed point
$h=0$, $t=0$ for most of the range of $l$. Therefore, the flow of the
Ising temperature field $t_{\rm I}$ is determined by the exponent
$y_{t_{\rm I}}= 7/8$ at $g_{\rm R}=9/4$ or $t=0$. Thus, at $g_{\rm R}=3$
it reaches the value $t_{\rm I}=b^{7/8}\delta t_{\rm I}=
 \exp(7\pi/4\sqrt{2 \delta h})\delta t_{\rm I}$. Since $h=-t=1/2$ up to
unimportant corrections, Eq.~(\ref{tovb}) leads to
\begin{equation}
\delta t_{\rm I} =  \frac{\beta}{2} \exp
 \left(-\frac{7\pi}{4\sqrt{2 \delta h}}\right)
\label{tfuh}
\end{equation}
which solves $\delta t_{\rm I}$ for all $\delta h>0$. This implies that
the Potts critical sheet connects to the KT point.
The resulting renormalization flow is sketched in Fig.~\ref{fig2}. 
Substitution of Eq.~(\ref{defp}) and $H=H_{\rm KT}+\delta H$ leads in
lowest order to
\begin{equation}
K \simeq \frac{1}{2} \left[ \ln \frac {\beta}{2} -
  \frac{7\pi}{4\sqrt{2 \alpha (H-H_{\rm KT})}} \right]
\label{shape}
\end{equation}
where $\alpha$ is a function of the $\alpha_j$. This equation determines
the shape of the critical line near the KT point of the model (\ref{Ising})
and is clearly incompatible with the prediction of the
invariant-polynomial scenario, Eq.~(\ref{eq:ktasympto2}).

\section{Numerical methods}
\label{sec4}
\subsection{Transfer-matrix calculations}
Most of the the transfer-matrix calculations were performed for $T>0$
so that we had to use a binary representation for the Ising spins,
leading to a transfer matrix of size $2^L \times 2^L$ for a system
with finite size $L$.
We define the spin lattice on the surface of a cylinder, whose axis
determines the transfer direction. We have used two choices for the
orientation of the lattice: one set of bonds parallel or perpendicular
with respect to the axis. For the first case one may apply a decoration
transformation to one half of the parallel bonds  in order to construct
a symmetric transfer matrix. However, the decoration of antiferromagnetic
bonds leads to complex weights which we wish to avoid. We have thus
used a non-symmetric transfer matrix, in combination with a suitable
tridiagonalization method to find the leading eigenvalues. These were
obtained for even linear system sizes up to $L=22$, which corresponds
with an actual finite size of $11 \sqrt{3}$ nearest-neighbor bonds.
The second construction, with a set of edges perpendicular to the
transfer direction, leads to a symmetric matrix when two layers of
spins are added. This allows the use of the conjugate-gradient method
which is, in our applications, more stable than the tridiagonalization
method. Finite-size calculations with $L$ multiples of 3 up to $L=24$
were performed using this second construction.

A sparse-matrix decomposition was used for both constructions. Most
of the technique is already implicit in the work of
Nightingale \cite{PN}. Further details are listed in Ref.~\cite{BWW}
which concerns the case of the honeycomb lattice, but the essential
steps are applicable to the triangular lattice as well.
During the analysis of the results of both types of transfer matrix
we found that they were mutually consistent. Furthermore it
became clear that the second transfer matrix, with a set of bonds
perpendicular to the transfer direction, allowed a somewhat more
accurate analysis. In the following we describe the situation of the
second construction.

For $T=0$ the transfer matrix decomposes in
a number of diagonal submatrices characterized by a conserved number
of `strings' so that the numerical diagonalization task simplifies.
The transfer-matrix construction for this case has been outlined
in Ref.~\cite{BN} and enabled the study of systems with linear sizes
up to $L=27$.

The magnetic correlation function along the coordinate $r$ in the
length direction of the cylinder is defined as
$g_{m}(r) = \langle s_{0}s_{r}\rangle$.
At large $r$, this correlation function decays exponentially with a
characteristic length scale $\xi$ that depends on $K$, $H$ and $L$
\begin{equation}
g_{m}(r) \propto {\rm e}^{-r/\xi(K,H,L)}
\end{equation}
and can be calculated from the largest two
eigenvalues $\lambda_0$ and $\lambda_1$ of the transfer matrix:
\begin{equation}
\xi^{-1}(K,H,L) =
\frac{1}{\sqrt{3}} \ln(\lambda_0/\lambda_1)
\end{equation}
where the factor $\sqrt{3}$ is a geometric factor, \ie  the 
ratio between the thickness of two layers added by the transfer
matrix and the length of a nearest-neighbor bond.
The significance of these relations lies in the fact that the assumption
of conformal invariance \cite{Cardyxi} links $\xi$ on the cylinder with
the magnetic scaling dimension $X_m$ (one half of the magnetic correlation
function exponent $\eta$). In terms of the scaled gap 
\begin{equation}
X_{m}(K,H,L) \equiv \frac{L}{2 \pi \xi(K,H,L)} 
\end{equation}
one has $X_{m}(K,H,L)\simeq X_{m}$ in the limit of large $L$.
Since the three-state Potts universal value of the magnetic scaling
dimension is known to be $X_{m}= \frac{2}{15}$, and the transfer-matrix
algorithm evaluates $X_m$ as a function of its arguments, one 
can find a numerical approximation to the critical value
of $K$ for a given value of $H$ or vice versa.
The  shape of the critical line prescribes 
the use of different ways in different regions. 
For small $H$ and large $|K|$, the critical line is 
almost parallel to the zero-field line, so that it becomes more
efficient to solve for $H$ than for $K$.

As a consequence of corrections to scaling, the solution will not
precisely coincide with the critical point. The effects of  
an irrelevant scaling field $u$ and a small deviation $t$ with
respect to the critical value of $H$ or $K$ are expressed by
\begin{equation}
X_{m}(K,H,L) =  X_m + au L^{y_i} + b t L^{y_t} + \cdots
\label{xhs}
\end{equation}
where $a$ and $b$ are unknown constants, $X_{m}= \frac{2}{15}$, 
$y_i=-\frac{4}{5}$ and $y_t=\frac{6}{5}$ for the 3-state Potts
universality class. Thus the solution for $K$ of
\begin{equation}
X_{m}(K,H,L) =\frac{2}{15}
\label{xhp}
\end{equation}
which is denoted ${K_{c}}^{(1)}(H,L)$, 
depends on  the finite size $L$ and the irrelevant field  as
\begin{equation} 
{K_{c}}^{(1)}(H,L)=K_c + c_1 L^{y_i-y_t} + \cdots
\end{equation}
because the two correction terms in Eq.~(\ref{xhs}) must cancel and
$t \propto {K_{c}}^{(1)}(H,L)-K_c $.
We thus generated sequences of iterated estimates of $K_c$ by solving 
${K_{c}}^{(2)}(H,L)$ and $c_1(L)$ in the equations 
\begin{equation} 
{K_{c}}^{(2)}(H,L)={K_{c}}^{(1)}(H,l) + c_1(L) l^{y_i-y_t} 
\end{equation}
for $l=L$ and $L+1$. These sequences appear to converge faster with
increasing $L$ than the ${K_{c}}^{(1)}(H,L)$. Remaining corrections
may be due to additional contributions to  Eq.~(\ref{xhs}), for instance
scaling as $ L^{2y_i}$.
Thus we defined ${K_{c}}^{3}(H,L)$ by solving the equation 
\begin{equation}
{K_{c}}^{3}(H,L) = {K_{c}}^{(2)}(H,l) + c_2 (L) l^{2 y_i-y_t} 
\label{Kc}
\end{equation}
for $l=L$ and $L+1$. Further estimates can be obtained with correction
exponents $2 y_i-2 y_t$, or by treating the correction exponents as a
free variable in which case three values of $l$ have to be used.
Several variations of this procedure were tried which leads to some
insight in the numerical inaccuracies of the fitting procedure.

Our final estimates of the critical points for $T>0$ are listed in
Tab.~\ref{tab:critpts}. The apparent accuracy of the critical points
is satisfactory for most of the field range, but it deteriorates rapidly
at small fields. Nevertheless this region has our special interest:
we wish to determine how the Potts line connects to the KT point on
the  $T=0$ line, because this is where the theoretical predictions
(see Sections~\ref{sec2} and \ref{sec3}) are markedly different.

We have also reconsidered the determination of the KT point at $T=0$
given in Ref.~\cite{BN}. In that work, the finite-size data for the
critical field $H_{\rm KT}$ were obtained by requiring that the 
scaled gap associated with the spin waves of period 6, \ie electric
charges $e=\pm 1$ were equal to the expected value $X_{1,0}=2/9$ at
the KT transition. 
These estimates $H_{\rm KT}(L)$ of the KT point, which were obtained
for system sizes up to $L=27$, were found to be considerably
size-dependent. They were fitted according to
$H_{\rm KT}(L) = H_{\rm KT} + a/(b+ \ln L) + c L^{-2}$
which led to extrapolated estimates $H_{\rm KT}$ that displayed only
a remarkably small size-dependence. On this basis, the final estimate
was given as $H_{\rm KT} = 0.266 \pm 0.010$ in Ref.~\cite{BN}.
In our present work we have reproduced these data for $H_{\rm KT}(L)$.   
We have not used the procedure that requires that
the scaled gap is equal for two subsequent system sizes. Since all 
points in the range $H \leq H_{\rm KT}$ satisfy this scaling equation
asymptotically, the solutions may converge to any point in this range,
depending on the corrections to scaling. This is another reason behind
the discrepancy mentioned in Sec.~\ref{int} concerning an earlier
result for the location of the KT point~\cite{QP}.
We have used the $H_{\rm KT}(L)$ data as input for several other
iterated fit procedures
consisting of subsequent extrapolation steps according to $L^{-2}$
behavior, and powers of $1/\ln L$. These fits led to results for
 $H_{\rm KT}$ that were rather consistently close to 0.26, with
differences up to 0.02.

As an independent approach we have estimated $H_{\rm KT}$ from the
requirement that the scaled gap based on the magnetic dimension
$X_{0,\frac{1}{3}}$ is equal to the expected value $1/8$ at $g_{\rm R}=9/4$.
Since such fractional magnetic charges (corresponding with vortices
of strength 2) do not exist in this model, this scaled
gap cannot directly be calculated for a fixed system size $L$.
However, it can be obtained by combining free-energy data for system
sizes $L= 3n \pm 1$, where $n$ is an integer, as explained in
Ref.~\cite{BN}. The same extrapolation procedures as above were
tried, and led to results again consistent with $H_{\rm KT} = 0.26$,
but with differences up to about 0.04.
Our final conclusion is $H_{\rm KT}=0.26 \pm 0.02$, similar to the
value presented in Ref.~\cite{BN} but with a slightly more conservative
error estimate.

\begin{figure}
\includegraphics[angle=0, width=7.5cm]{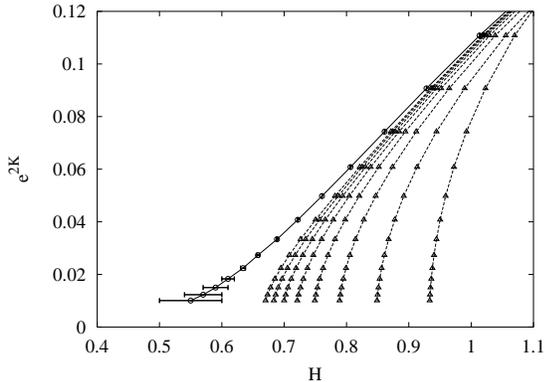}
\caption[xxx]
{Finite-size solutions for the critical points, and our final estimates
in the region of small field and temperature. The dashed lines connect
the solutions shown as triangles ($\triangle$). From right to left the lines
show data for finite sizes $L=3$, 6, $\cdots$, 24. The solid line with
circles $(\circ)$ indicates our final estimated result for the critical
line.}
\label{fig3}
\end{figure}

The numerical results for the critical points are combined in
Fig.~\ref{fig1}, the phase diagram in the $(H,{\rm e}^{2K})$ plane.
\begin{table}
\caption{\label{tab:critpts}
Extrapolated results for selected points on the critical line.
 }
\vskip 2mm
\begin{tabular}{|c|c|c|c|c|c|}
 \hline
$\# $ & $H$    &    $K$             &$\# $& $H$&      $K$       \\
 \hline
 1 &  0.55(5)  &   -2.3             & 20&  1.85&  -0.759438 (2) \\
 2 &  0.57(3)  &   -2.2             & 21&  1.90&  -0.755049 (1) \\
 3 &  0.59(2)  &   -2.1             & 22&  1.95&  -0.751498 (2) \\
 4 &  0.610(10)&   -2.0             & 23&  2.0 &  -0.748715 (2) \\
 5 &  0.634(4) &   -1.9             & 24&  2.1 &  -0.745199 (1) \\
 6 &  0.658(2) &   -1.8             & 25&  2.178& -0.744130 (1) \\
 7 &  0.6885(10)&  -1.7             & 26&  2.3 &  -0.744958 (1) \\
 8 &  0.7219(3)&   -1.6             & 27&  2.4 &  -0.747586 (1) \\
 9 &  0.7607(1)&   -1.5             & 28&  2.5 &  -0.751708 (1) \\
10 &  0.8      &   -1.414    (1)    & 29&  2.75&  -0.7673233(4) \\
11 &  0.9      &   -1.2395   (2)    & 30&  3.0 &  -0.7887774(4) \\
12 &  1.0      &   -1.11422  (1)    & 31&  3.25&  -0.8145143(3) \\
13 &  1.1      &   -1.02100  (1)    & 32&  3.5 &  -0.8434661(2) \\
14 &  1.2      &   -0.95030  (1)    & 33&  4.0 &  -0.9082113(2) \\
15 &  1.3      &   -0.896040 (5)    & 34&  4.5 &  -0.9791030(2) \\
16 &  1.4      &   -0.854175 (2)    & 35&  5.0 &  -1.0539340(2) \\
17 &  1.5      &   -0.821890 (1)    & 36&  5.5 &  -1.1313743(2) \\
18 &  1.6      &   -0.797164 (2)    & 37&  6.0 &  -1.2105830(3) \\
19 &  1.75     &   -0.771064 (3)    &   &      &                \\
 \hline
\end{tabular}
\end{table}
We remark that for large $|K|$ and relatively small $H$, 
the solutions for $H$ become strongly  finite-size dependent and
slowly convergent. This problem is apparently due to 
the proximity of the KT transition
at $K=-\infty$.  This is illustrated in Fig.~\ref{fig3}, in which a
set of lines represent the finite-size solutions for 
$L=3$, 6, $\cdots$, 24 together with the extrapolated critical line.
On the basis of our limited range of finite system sizes, the
estimation of the critical points thus becomes increasingly difficult
for large $|K|$.

In order to provide further justification for our assumption that the
critical line belongs to the 3-state Potts universality class (used
for the determination of the critical points) we perform a consistency
test by calculating the conformal anomaly $c$ at the estimated
critical points at $H=1.5$, 2.5, 3.5 and 4.5. Iterated fits  similar to
those used for the calculation of the critical points were applied.
All these results are consistent with the exact value $c=\frac{4}{5}$.
The error margin varies between a few times $10^{-3}$ for $H=1.5$
and a few times $10^{-5}$ for $H=3.5$.
In comparison with previous work \cite{NK}, these results further
restrict the scale of possible deviations from 3-state Potts universality.

\subsection{Monte Carlo results}  
\label{mcres}
The apparent difficulty to obtain accurate critical points for small $H$
by the transfer-matrix method  invites further investigation
by means of Monte Carlo simulations, which allow the use of much larger
finite sizes. In particular we determine how the critical amplitudes
behave for small $T$, and make a comparison with the renormalization
prediction.
To this purpose we define a specific-heat-like quantity $C$, \ie
the second order derivative of the free energy to a parameter
conjugate to an energy-like density in the Hamiltonian, for which we
may take the magnetization. Indeed the Ising field $H$ drives the
Potts-like transition to the ordered state (except at the maximum of the
$T$ vs. $H$ curve) and thus plays the role of the temperature in the
Potts model. We thus define $C$ by
\begin{equation}
 C = \frac{\partial^2 f} {\partial H^2}= 
N (\langle m^2 \rangle - \langle m \rangle^2)
\label{sph}
\end{equation}
where $m$ is the Ising magnetization.  Similarly, we define a quantity
similar to the magnetic
susceptibility of the Potts model. In the $q=3$ Potts model, the
zero-field magnetic susceptibility can be expressed in magnetization
fluctuations by
\begin{equation}
 \chi = N \langle m_{\rm P}^2 \rangle 
\label{chp}
\end{equation}
where $m_{\rm P}^2=n_1^2+n_2^2+n_3^2-n_1n_2-n_2n_3-n_3n_1$ expresses
the Potts magnetization in terms of the densities $n_i$ of Potts
variables in state $i$. In the scaling context of the present scaling
analysis, the densities $n_i$ may be defined
as the number of minus-spins on sublattice $i$. Thus $\chi$ describes
the response of the model to staggered fields acting on the Ising spins.

\begin{figure}
\includegraphics[angle=0, width=7.5cm]{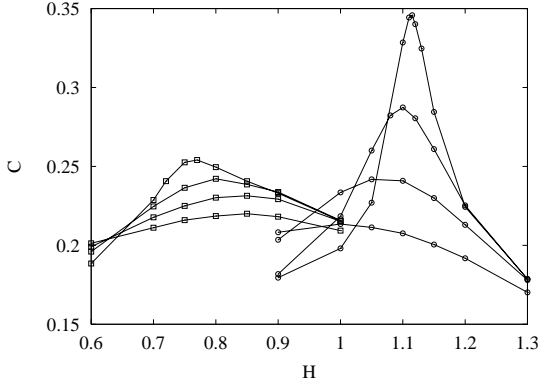}
\caption[xxx]
{Specific heat-like quantity $C$ versus field $H$ for two values of
the Ising coupling, $K=-1.0$ ($\circ$) and $-1.5$ ({\tiny$\square$}).
Both sets of lines display data for system sizes $L=12$, 24, 48, and 96.
These two cases indicate that the amplitude of the divergence of $C$ 
decreases with decreasing field $H$.  }
\label{fig4}
\end{figure}

\begin{figure}
\includegraphics[angle=0, width=7.5cm]{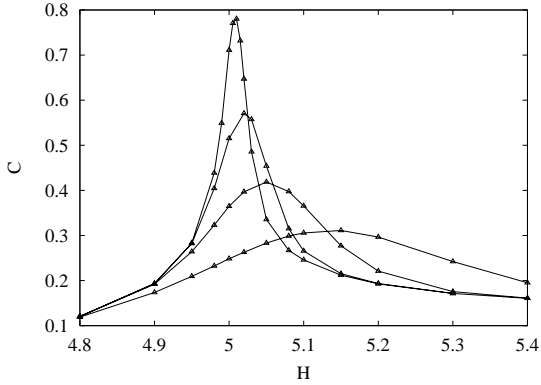}
\caption[xxx]
{Specific heat-like quantity $C$ versus $H$ at Ising coupling
$K=-1.053934$, near the critical point at $H=5.0$.  In comparison with 
Fig.~\ref{fig4}, the finite-size divergence is much stronger.}
\label{fig5}
\end{figure}

\begin{figure}
\includegraphics[angle=0, width=7.5cm]{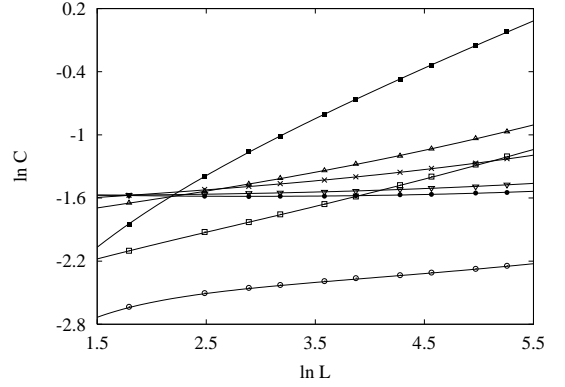}
\caption[xxx]
{Specific heat-like quantity $C$ versus system size on a double
logarithmic scale.
Data are shown for system sizes $6\leq L \leq 192$ at seven points
$(H,K)$ on the critical line. The symbol $\bullet$ represents
$H=0.61$; $\triangledown$: $H=0.658$; $\times$: $H=0.8$; $\triangle$:
$H=1.0$; {\tiny$\square$}: $H=1.5$;
{\large $\circ$}: $H=2.0$; {\tiny$\blacksquare$}: $H=5.0$.
The seven lines represent the fitted results.
The statistical errors are not shown in this figure. They
do not exceed the thickness of the lines.}
\label{fig9}
\end{figure}

\begin{figure}
\includegraphics[angle=0, width=7.5cm]{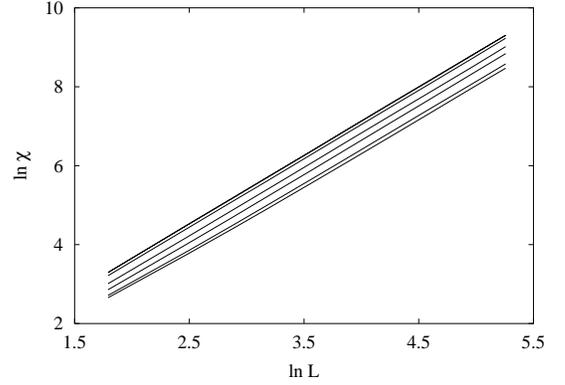}
\caption[xxx]
{Susceptibility-like quantity $\chi$ versus system size
$L(=6,\cdots,192)$ on a double 
logarithmic scale. Data are shown for seven points $(H,K)$ on the critical
line. From bottom to top: $H=0.61$, 0.658, 0.8, 1.0, 1.5, 2.0, and 5.0 
respectively. The two lines for $H=2.0$ and 5.0 coincide on this scale. 
The statistical errors are not shown in this figure. They
do not exceed the thickness of the lines.  
}
\label{fig10}
\end{figure}

The simulations used triangular $L \times L$ lattices with periodic
boundary conditions. We used a combination of the standard Metropolis
algorithm and the geometric cluster method \cite{JHHB}. The latter method
executes nonlocal updates and leads to a faster relaxation. But it does
not change the Ising magnetization. For this reason also Metropolis steps
were included.  First we sampled $C$ in a suitable range of $H$, to
study its divergence, at fixed values of coupling $K$
which are taken from Table~\ref{tab:critpts}.  The results are shown in
Figs.~\ref{fig4} and \ref{fig5} and show that the finite-size divergence
of $C$ at the critical line becomes weaker when $H$ decreases.
In order to study this phenomenon in a more quantitative sense,
we have determined $C$ and $\chi$ at several
critical points taken from Table~\ref{tab:critpts} for several system
sizes.  Results for $H=0.61$, 0.658, 0.8, 1.0, 1.5, 2.0 and 5.0 are shown in
Fig.~\ref{fig9}  and Fig.~\ref{fig10}.

Finite-size scaling of the free energy density of a system with finite
size $L$ can be expressed by
\begin{equation}
 f(t,h,u,L) = L^{-d} f(L^{y_{t}}t,L^{y_{h}}h,L^{y_{i}} u,1)+g(t,h,u)
\label{fre}
\end{equation}
where $t$, $h$, and $u$ denote the temperature, magnetic field and 
irrelevant field respectively, and $g$ the regular part of the 
transformation. Differentiation of $f$ yields the scaling behavior of
the quantities $C$ and $\chi$ as
\begin{displaymath}
 C(u,L) =C_0+L^{2y_{t}-d} C(L^{y_{i}} u,1)
\end{displaymath}
\begin{equation}
    =C_0+L^{2y_{t}-d} (b_0+b_1 L^{y_{i}} u+b_2 L^{2y_{i}} u^2 + \cdots)
\label{sphc}
\end{equation}
and
\begin{displaymath}
 \chi(u,L) =\chi_0+L^{2y_{h}-d} \chi (L^{y_{i}} u,1)
\end{displaymath}
\begin{equation}
  =\chi_0+L^{2y_{h}-d} (b_0+b_1 L^{y_{i}} u+b_2 L^{2y_{i}} u^2 + \cdots)
\label{susc}
\end{equation}
We have fitted the numerical data by these two equations 
and thus derived the amplitudes listed in Tab.~\ref{table1} and 
Tab.~\ref{table2}. The amplitude $b_0$  of the leading divergence of $C$
decreases with field except close to the maximum of the critical line
in the $(H,T)$ plane. At the maximum the field $H$ fails to bring the
system into the ordered phase and the amplitude $b_0$ thus vanishes.
Also for the susceptibility-like quantity $\chi$ the amplitude $b_0$ of
the finite-size divergence, shown in Tab.~\ref{table2}, decreases
regularly when the KT point is approached.
\begin{table}                                                              
\caption{\label{table1} 
Parameters describing the finite-size behavior of the specific heat-like
quantity $C$.
The third column is the amplitude $b_0$ of the leading divergent term.
The amplitudes for $H=1.5, 2.0$ are relatively small because the
critical line runs almost parallel to the field direction.
The column corresponding with $b_i$ $(i=1,2)$ are the irrelevant corrections
amplitudes.
}
 
\vskip 2mm                                                                
\begin{tabular}{|c|c|c|c|c|}  
\hline
 $H$    & $C_0$     &   $b_0$      &  $b_1$    & $b_2$      \\
\hline                                                                      
 0.61   &  0.188(2) &  0.0026(2)   & 0.027(3)  &   -        \\
 0.658  &  0.196(2) &  0.0039(3)   & 0.007(3)  &   -        \\
 0.8    &  0.192(3) &  0.0126(4)   & -0.022(4) &   -        \\
 1.0    &  0.151(3) &  0.0285(6)   & -0.035(5) &   -        \\ 
 1.5    &  0.097(2) &  0.0254(4)   & -0.055(3) &   -        \\
 2.0    &  0.084(3) &  0.0029(3)   & -0.015(6) &   -0.097(2)\\
 5.0    & -0.096(6) &  0.131(2)    & -0.031(9) &   -        \\
\hline
\end{tabular}
\end{table}                                                         

\begin{table}
\caption{\label{table2}
Parameters describing the finite-size behavior of the susceptibility-like
quantity $\chi$.
The third column is the amplitude $b_0$ of the leading divergent term.
It decreases regularly with the field. The next columns shows the
irrelevant correction amplitudes $b_i$ $(i=1,2)$.}
\vskip 2mm
\begin{tabular}{|c|c|c|c|c|}
\hline
 $H$   &  $\chi_0$    &    $b_0$   &      $b_1$   & $b_2$    \\
\hline
 0.61  & -17.5(18)    & 0.5281(12) &   -0.143(50) & 16.3(16) \\
 0.658 & -10.5(20)    & 0.5896(14) &   -0.142(55) & 2.88(36) \\
 0.8   & -18.7(37)    & 0.7675(24) &   -0.75(10)  & 18.1(33) \\
 1.0   &   0.95(11)   & 0.9046(17) &   -0.164(25) &  -       \\
 1.5   &   0.41(10)   & 1.1209(12) &   -0.103(24) &  -       \\
 2.0   &   0.10(14)   & 1.2130(24) &   -0.003(34) &  -       \\
 5.0   &  -0.51(14)   & 1.2133(22) &    0.070(37) &  -       \\
\hline
\end{tabular}
\end{table}

The behavior of the amplitude $b_0$ at small field and low temperature
follows from the  renormalization-flow analysis. 
Starting from a point in the vicinity of the KT point, we renormalize 
until we arrive at the boundary with the region dominated by the 3-state
Potts fixed point.  Let $b_{\rm KT}$ be the corresponding scale factor.
Since the specific heat-like quantity $C$ is defined by means of 
differentiation of the free energy to the uniform field $H$, we keep
track of how $H$ changes under this transformation. The marginality of
$\delta h$ at $g_R=9/4$ is expressed by Eq.~(\ref{tra2p}): when we 
write $h=t+\delta h$, it is clear that $\delta h$ varies only by a factor
of order 1 as long as $t$ is of order 1. In the context of scaling, we
thus have $\delta h' \approx \delta h$ where the prime indicates the
value at the boundary.
Within the Potts region we rescale the system to size 1 with the
remaining scale factor $L/b_{\rm KT}$    
\begin{equation}
\delta h{''}=\left(\frac{L}{b_{\rm KT}}\right)^{\frac{6}{5}} \delta h{'}
\approx L^{\frac{6}{5}} b_{\rm KT}^{-\frac{6}{5}} \delta h
\end{equation}
where the Potts temperature exponent $\frac{6}{5}$ applies because it
corresponds with the Ising magnetic field.
The behavior of $C$ follows as
\begin{equation}
 C = \frac{\partial^2 f} {\partial h^2}
=L^{-d} \frac{\partial^2}{\partial h^2} 
f(L^{\frac{6}{5}} b_{\rm KT}^{-\frac{6}{5}} h,1)
=L^{\frac{2}{5}} b_{\rm KT}^{-\frac{12}{5}} f{''} + {\rm const}
\label{b0c}
\end{equation}

The susceptibility-like quantity $\chi$ is obtained by differentiation of
$f$ to the staggered Ising field, which, as explained in Ref.~\cite{N}, is
associated with a Gaussian spin wave perturbation of period $p=6$, \ie
with electric charges $e=\pm 1$. The exponent of the staggered field thus
takes the value $2-X_{1,0}=16/9$ at the KT fixed point.  Therefore, at
the boundary with the Potts region we have
\begin{equation}
h_{\rm st}^{'}=b_{\rm KT}^{\frac{16}{9}} h_{\rm st}
\end{equation}   
Within the region dominated by the Potts fixed point, the magnetic
exponent $y_{m}=28/15$ applies. Renormalization
with the remaining scale factor $L/b_{\rm KT}$ leads to
\begin{equation}
h_{\rm st}^{''}=\left(\frac{L}{b_{\rm KT}}\right)^{\frac{28}{15}} 
h_{\rm st}^{'} =L^{\frac{28}{15}} b_{\rm KT}^{-\frac{4}{45}} h_{\rm st}
\end{equation}
so that $\chi$ scales as
\begin{displaymath}
 \chi = \frac{\partial^2 f} {\partial h_{\rm st}^2}
=L^{-d} \frac{\partial^2}{\partial h_{\rm st}^2} 
f(L^{\frac{28}{15}} b_{\rm KT}^{-\frac{4}{45}} h_{\rm st},1) =
\hspace{20mm}
\end{displaymath}
\vspace{-5mm}
\begin{equation}
\hspace{45mm}
L^{\frac{26}{15}} b_{\rm KT}^{-\frac{8}{45}} f{''} + {\rm const}
\label{b0x}
\end{equation}
According to Sec.~\ref{rgflow}, rescaling by  a factor $b_{\rm KT}$
results in an Ising temperature field
$b_{\rm KT}^{\frac{7}{8}} {\rm e}^{2K}=\beta/2$ so that
$b_{\rm KT} \propto {\rm e}^{-\frac{16}{7}K}$. 
For strong coupling the renormalization scale $b_{\rm KT}$ is large, 
which is indicative of the crossover phenomenon close to the KT 
transition. There we need large system sizes $L>b_{\rm KT}$ in order
to reach the vicinity of the Potts fixed point.  The substitution of
$b_{\rm KT}$  into Eq.~(\ref{b0c}) and Eq.~(\ref{b0x}), 
leads to
\begin{equation}
C = L^{\frac{2}{5}}{\rm e}^{\frac{192}{35} K } f{''} + {\rm const}
\label{b0cc}
\end{equation}
and
\begin{equation}
\chi =L^{\frac{26}{15}}{\rm e}^{\frac{128}{315} K } f{''} + {\rm const}
\label{b0xx}
\end{equation}
{}From a comparison of Eqs.\,(\ref{sphc}) and \,(\ref{b0cc}), 
and of Eqs.\,(\ref{susc}) and \,(\ref{b0xx}), we expect that
$b_0 \propto {\rm e}^{\frac{192}{35} K }$ for $C$, and 
$b_0 \propto {\rm e}^{\frac{126}{315} K }$ for $\chi$, when $|K|$
is large enough.
We thus expect a linear relation between $ \ln b_0$ and $K$ for
sufficiently strong coupling $K$.   
A fit to the numerical data  yields the slopes as about 2.8(1) for $C$ 
and 0.64(4) for $\chi$.
These spoles do not agree accurately with the analytic values.
This suggests that the Ising temperature used for the calculation
of the amplitudes is not small enough. However, the qualitative
amplitude dependence is reproduced, and the rough agreement
suggests that we are not far away from the asymptotic regime.

\section{Fit and discussion}
\label{sec5}
\subsection{Roots of the O(2) invariant polynomials}
\begin{figure}
\includegraphics[angle=0, width=7.5cm]{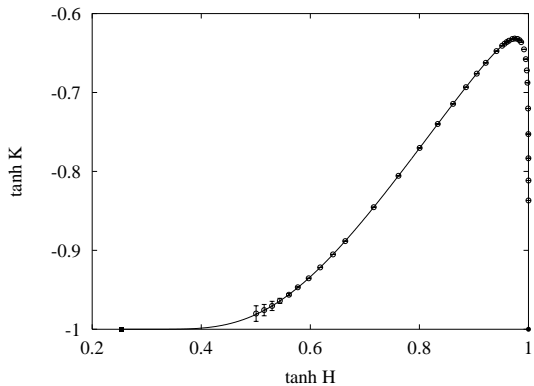}
\caption[xxx]
{Phase diagram in the $(h,z)$ plane (where $h=\tanh H$ and $z=\tanh K$).
The KT point is denoted as {\tiny $\blacksquare$}, and the LG point,
which is Baxter's hard-hexagon model, as $\bullet$.}
\label{fig14}
\end{figure}

The number of 37 critical points in Table~\ref{tab:critpts} together with
the additional KT point $( h, z ) = ( h_{\tmop{KT}}, - 1 )$ is sufficient
to attempt fits of O(2) invariant polynomials \ref{eq:f2ek} up to order 10.
We have not used data points 1 and 2 because of their limited accuracy,
and performed a least-squares fit to the remaining 35 data points for
$z>-1$. We have also tried direct fits to several subsets of these.
In each case we have investigated the effect of enforcing the curve
to pass through the KT point, or to extract the value of
$h_{\tmop{KT}}$ from the fit. It is found that the
coefficients in the equation $f_4 = 0$ are not flexible enough to even
qualitatively fit the numerical data. The least-squares fit to $f_6 = 0$
excluding the KT point consists of two avoiding solutions,
which lead to 2 disconnected `critical' lines
which have unphysical ranges. For $z=-1$ one line
terminates at $h^{\ast} \approx 0.501 > h_{\tmop{KT}}$. When enforcing
the KT point at $h_{\tmop{KT}} = 0.25$, the two avoiding branches repel
one another even stronger. Direct fits to different subsets of critical
points lead to similar results.
Fits to higher order equations $f_8 = 0$ and $f_{10} = 0$ display the same
problems. Even more avoiding solutions enter. The numerical problems 
are clearly displayed by the values of the fitted coefficients, which span 
a range of many orders of magnitude.  In summary,
the roots of invariant polynomial equations cannot fit the critical
curve. The main problem is the approach of the curve to the KT limit
imposed by Eq.~(\ref{eq:ktasympto2}): all such
roots approach the KT point vertically in the $(h,z)$ plane, whereas the
numerical data in Fig.~\ref{fig14} indicate a horizontal approach.

Sufficiently far away from the KT point, the problem dissolves, and our
numerical data for the critical points can well be approximated by means
of invariant polynomials.  For example, the polynomial of order 6
can reproduce the critical points for $H \geq 1.5$ within the error
margins quoted in Table~\ref{tab:critpts}. The coefficients,
determined by means of a least-squares fit, are listed in Tab.~\ref{o2p}.

\begin{table}
\caption{\label{o2p}
Coefficients $\kappa_i^j$  $(i=0,...,6; j=1,2,3)$ of the invariant
polynomial $f_6$, Eq.~(\ref{eq:f2ek}). The condition for criticality
reads $f_6=0$.
  }
\vskip 2mm
\begin{tabular}{|c|c|c|}
\hline                                                        
  $\kappa_0^1$    &  $\kappa_1^1$     &    $\kappa_2^1$  \\      
\hline                                                        
 -0.002942307242  &  -0.097656582607  &   2.725303204552 \\
\hline                                                        
\hline                                                        
  $\kappa_0^2$    &    $\kappa_1^2$   &   $\kappa_2^2$   \\     
\hline                                                        
 -0.000000000017  &   0.000351055639  &  -0.020534345140 \\
\hline                                                        
\hline                                                        
  $\kappa_3^2$    &    $\kappa_4^2$   &   $\kappa_3^3$   \\       
\hline                                                        
  0.044511873969  &   0.660086779714  &   0.000001077354 \\          
\hline                                                        
\hline                                                        
  $\kappa_4^3$    &    $\kappa_5^3$   &   $\kappa_6^3$   \\     
\hline                                                        
 -0.002865984972  &  -0.084578467922  &   1.000000000000 \\          
\hline                                                        
\end{tabular}                                                 
\end{table} 

\subsection{The renormalization solution for small field}
\begin{figure}
\includegraphics[angle=0, width=7.5cm]{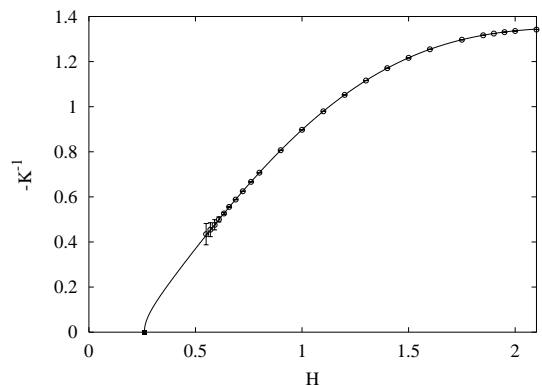}
\caption[xxx]
{Phase diagram in the ($H,K^{-1}$) plane. The symbol $\circ$
denotes the Potts transition points; {\tiny $\blacksquare$} the KT point;
and the solid line describes the fit of the expression based on the 
renormalization prediction for the
critical line in the small field region.}
\label{fig12}
\end{figure}

For small field we expand Eq.\,(\ref{shape}) and take into account 
higher order terms in the physical fields. This leads to
\begin{equation}
 -\frac{1}{K}= \sum_{j=1,2,\cdots} a_j (H-H_{\rm KT})^{j/2} 
\label{TH1}
\end{equation}
The numerical data for the critical points for $H \leq 1.75$ are
fitted satisfactorily (\ie within the error margins quoted in
Table~\ref{tab:critpts}) by this formula using $6$
coefficients. The numerical results and the fitted function are shown
in Fig.~\ref{fig12}. The values of the coefficients are listed in
Tab.~\ref{rgf}.

\begin{table}
\caption{\label{rgf}
Coefficients $a_j$ of the expansion (\ref{TH1}) of $-K^{-1}$ in 
powers of $(H-H_{\rm KT})^\frac{j}{2}$.
  }
\vskip 2mm
\begin{tabular}{|c|c|c|}
\hline
   $a_1$    &    $a_2$  &   $a_3$  \\ 
\hline
0.487432    & 0.119116  & 0.765066 \\
\hline
\hline
  $a_4$     & $a_5$     & $a_6$    \\
\hline
1.017104    &-1.949253  & 0.652161 \\   
\hline
\end{tabular}
\end{table}

\section{Conclusion}
\label{con}
The invariant-polynomial scenario formulated in Sec.~\ref{sec2} and
the renormalization scenario formulated in Sec.~\ref{sec3} lead to
analytic expressions for the critical line in the $(H,T)$ diagram that
are mutually inconsistent for $T \downarrow 0$ at finite $H$. This
shows that at least one of the underlying assumptions must be incorrect. 
The renormalization prediction appears to successfully describe the
numerical data for small $|H|$. Although the asymptotic regime is not
quite reached (as can, for instance, be seen in Fig.~\ref{fig12} where
the leftmost points behave almost linearly instead of as a square root),
an asymptotic expansion leads to an accurate description of the data,
and allows a smooth extrapolation to zero Ising temperature in
agreement with Eq.~(\ref{shape}). The analysis in Sec.~\ref{mcres}
of the critical amplitudes fits precisely in this picture.
 Thus our analysis does not give reasons to doubt that 
the renormalization scenario correctly describes 
the essential physics of the model near the KT transition.

In contrast, the invariant-polynomial scenario does not agree with
the numerical data. It predicts a `vertical' approach to the KT
point in the $(h,z)$ diagram (see Fig.~\ref{fig14}) where it should
be horizontal.  Our interpretation is that the assumption of
analyticity of the critical line in the $(h,z)$ parametrization is
false at the KT point, so that the line cannot be described by the
zeroes of a polynomial of a finite order. 

Since it now appears that the invariant-polynomial scenario fails
in the case of the triangular-lattice Ising model, the question arises
whether similar, apparently successful, analyses of the critical lines
of the honeycomb- and the square-lattice Ising model in terms of
invariant polynomials \cite{WWB,BW} have to be reconsidered.
Here we may point at the simpler topology of the $(H,T)$ diagram for
the honeycomb and the square lattices: the critical line connects to
$T=0$ only in the lattice-gas points $H= \pm \infty$. In the case of
the triangular lattice model, crossover phenomena near the KT point
are responsible for the nonanalytic `shape' of the  critical line.
In the absence of such crossover phenomena, there is no inconsistency
with the invariant-polynomial scenario, and our present analysis has
therefore no direct consequences for the work presented in
Refs.~\cite{WWB,BW}.

\begin{acknowledgments} We are indebted to Prof. Fa Y. Wu and Prof.
Bernard Nienhuis for several illuminating discussions and useful
conversations, to Dr. Xue-Ning  Wu for her valuable contributions in
the earlier stages of this project, and to Dr. Jouke R. Heringa for 
his contribution to the development of the geometric cluster algorithm
used in this work.
\end{acknowledgments}

\newpage

\begin{thebibliography}{widest-label}
\bibitem{Hout}
R.M.F. Houtappel, Physica {\bf 16}, 425 (1950).
\bibitem{ST}
J. Stephenson, J. Math. Phys. {\bf 11}, 413 (1970).
\bibitem{SA}
S. Alexander, Phys. Lett. A {\bf 54}, 353 (1975).
\bibitem{KS}
W. Kinzel and M. Schick, Phys. Rev. B {\bf 23}, 3435 (1981).
\bibitem{PN}
M.P. Nightingale, Proc. Kon. Ned. Ak. Wet. B {\bf 82}, 235 (1979).
\bibitem{NK}
J.D. Noh and D. Kim, Int. J. Mod. Phys. B {\bf 6}, 2913 (1992).
\bibitem{TS}
M.N. Tamashiro and S.R. Salinas, Phys. Rev. B. {\bf 56}, 8241 (1997).
\bibitem{ZR}
Z. R\`acz, Phys. Rev. B. {\bf 21}, 4012 (1980).
\bibitem{BaxterHH}
R.J. Baxter, {\it Exactly Solved Models in Statistical Physics},
(Academic Press 1982).
\bibitem{N}
B. Nienhuis, H.J. Hilhorst and H.W.J. Bl\"ote,
J. Phys. A {\bf 17}, 3559 (1984).
\bibitem{BH}
H.W.J. Bl\"ote and H.J. Hilhorst, J. Phys. A {\bf 15}, L631 (1982).
\bibitem{BN}
H.W.J. Bl\"ote and M.P. Nightingale, Phys. Rev. B {\bf 47}, 15046 (1993).
\bibitem{KT}
J.M. Kosterlitz and D.J. Thouless,
J. Phys. C {\bf 5} L124 (1973); J. Phys. C {\bf 6} 1181 (1973).
\bibitem{BNH}
H.W.J. Bl\"ote, M.P. Nightingale, X.N. Wu and A. Hoogland, Phys. Rev.
B {\bf 43}, 8751 (1991).
\bibitem{QP}
S.L.A. de Queiroz, T. Paiva, J.S. de S\'a Martins
and R.R. dos Santos, Phys. Rev. E. {\bf 59}, 2772 (1999).
\bibitem{Qp}
S.L.A. de Queiroz, private communication (2003). The extrapolated
slope $S_L$ in Table~I of the preceding reference is thus the inverse of
our $H_{\rm KT}$.
\bibitem{FYWu}
F.Y. Wu,  J. Math. Phys. (NY) {\bf 15}, 687 (1974).
\bibitem{WuWu}
F.Y. Wu and X.N. Wu, J. Stat. Phys. {\bf 90}, 41 (1988).
\bibitem{WWB}
F.Y. Wu, X.N. Wu and H.W.J. Bl\"ote, Phys. Rev. Lett. {\bf 62}, 2773 (1989).
\bibitem{BW}
H.W.J. Bl\"ote and X.N. Wu, J. Phys. A {\bf 23}, L627 (1990).
\bibitem{bib:perk}
J.H.H. Perk, F.Y. Wu and X.N. Wu, J. Phys. A {\bf 23}, L131 (1990).
\bibitem{JK}
J.V. Jos\'e, L.P. Kadanoff, S. Kirkpatrick and D.R. Nelson, Phys. Rev. B
 {\bf 16}, 1217 (1977).
\bibitem{BNDL}
B. Nienhuis, in {\it Phase Transitions and Critical Phenomena} Vol. 11,
eds. C. Domb and J.L. Lebowitz (Academic, London, 1987).
\bibitem{XH1}
X.F. Qian and H.W.J. Bl\"ote, unpublished (2003).
We have introduced next-nearest neighbor interactions $K_{\rm nnn}$ and
used a transfer-matrix method to locate the $g_{\rm R}=3$ point and the
approximate location of the Potts critical line for that value of
$K_{\rm nnn}$.
\bibitem{Landau}
D.P. Landau, Phys. Rev. B {\bf 27}, 5604 (1983)
\bibitem{BWW}
H.W.J. Bl\"ote, F.Y. Wu and X.N. Wu, Int. J. Mod. Phys. B {\bf 4}, 619 (1990).
\bibitem{Cardyxi}
 J.L. Cardy, J. Phys. A {\bf 17}, L358 (1984).
\bibitem{JHHB}
J.R. Heringa and H.W.J. Bl\"ote,  Phys. Rev. E {\bf 57}, 4976 (1998).
\end{thebibliography}
\end{document}